\documentclass[12pt]{article}
\usepackage{graphicx}
\usepackage{amsmath}
\usepackage{amssymb}

\begin{document}

\title{Lyapunov exponents in fundamental models of nonlinear resonance}

\author{Ivan~I.~Shevchenko$^{1,2}$}

\maketitle

\noindent $^1$Saint Petersburg State University, 7/9
Universitetskaya nab., \\ 199034 Saint Petersburg, Russia \\
email: {\tt i.shevchenko@spbu.ru}

\noindent $^2$Institute of Applied Astronomy, Russian Academy of
Sciences, \\ 191187 Saint~Petersburg, Russia

\begin{abstract}
The problem of analytical estimation of the Lyapunov exponents and
Lyapunov timescales of the motion in multiplets of interacting
nonlinear resonances is considered. To this end, we elaborate a
unified framework, based on the separatrix map theory, which
incorporates both an earlier approach for the first fundamental
model of perturbed resonance (given by the perturbed pendulum
Hamiltonian) and a new one for its second fundamental model (given
by the perturbed Andoyer Hamiltonian). Within this framework, new
accurate estimates for the Lyapunov timescales of the inner and
outer subsystems of the Solar planetary system are presented and
discussed.
\end{abstract}

\medskip

\medskip


\bigskip

\section{Introduction}

Estimations of the Lyapunov exponents provide insights in most
fundamental properties of dynamical systems
\cite{C79,LL92,Meiss92}. The maximum Lyapunov exponent $L$
characterizes the maximum exponential divergence rate of
initially-close trajectories in phase space.

The maximum Lyapunov exponent is defined by the formula (e.g.,
\cite{LL92}):

\begin {equation}
L = \limsup _ {{t \to \infty} \atop {d (t_0) \to 0}}
\frac{1}{t-t_0} \ln \frac{d (t)}{d (t_0)} , \label {def_lce}
\end {equation}

\noindent where $d(t_0)$ is the distance (in the phase space of
motion) between two nearby initial conditions for two trajectories
at the initial instant of time $t_0$, and $d(t)$ is the distance
between the evolved initial conditions at time $t$.

The so-called Lyapunov time (or timescale), $T_\mathrm{L} =
L^{-1}$, characterizes the time of predictable dynamics. The
behaviour of any dynamical system cannot be predicted on
timescales much greater than its $T_\mathrm{L}$; see reviews on
physical and astrophysical applications in
\cite{C79,LL92,A06,S20}.

In this article, we consider the problem of analytical estimation
of the Lyapunov exponents and Lyapunov timescales of the motion in
multiplets of interacting nonlinear resonances.

To describe the interaction of resonances in multiplets, we
consider two fundamental models: the perturbed pendulum model
(introduced in \cite{C79} as a ``universal'' one) and the
perturbed Andoyer model (introduced in \cite{HL83} to describe
resonances in orbital dynamics). Hereafter, we denote these two
fundamental models of the perturbed nonlinear resonance (the
guiding one in the multiplet) as FMR1 and FMR2, respectively.

We elaborate a unified framework, based on the separatrix map
theory, which incorporates both an earlier approach
\cite{S20,S02CR,S14PLA} for FMR1 (given by the perturbed pendulum
Hamiltonian) and a new one for FMR2 (given by the perturbed
Andoyer Hamiltonian).

Within this framework, we perform new accurate estimates for the
Lyapunov timescales of the inner and outer subsystems of the Solar
planetary system, and discuss their conformity with earlier
analytical and numerical-experimental estimates.

We assume that the resonances in the multiplet have comparable
strengthes. This choice is inspired by the fact that in realistic
applications the perturbations are usually not at all weak
(otherwise the chaotic component of phase space can be simply
unimportant); see examples in \cite{S20,S07IAU}.

\section{The first fundamental model of resonance}
\label{sect_FMR1}

First of all, let us briefly review relevant known results for
FMR1, necessary for the following considerations.

We define the first fundamental model of perturbed nonlinear
resonance (FMR1) by the following paradigmatic Hamiltonian
\cite{S14PLA,S99,S00JETP}:

\vspace{-3mm}

\begin{equation}
H = \frac{{\cal G} p^2}{2} - {\cal F} \cos \phi +
    a \cos(\phi - \tau) + b \cos(\phi + \tau).
\label{h}
\end{equation}

\noindent The first two terms in Eq.~(\ref{h}) represent the
Hamiltonian $H_0$ of the unperturbed pendulum:

\begin{equation}
H_0 = \frac{{\cal G} p^2}{2} - {\cal F} \cos \phi, \label{h0_FMR1}
\end{equation}

\noindent where $\phi$ is the pendulum angle (the resonance phase
angle), and $p$ is the conjugated momentum. The periodic
perturbations are given by the third and fourth terms in
formula~(\ref{h}); in them, $\tau$ is the phase angle of
perturbation: $\tau = \Omega t + \tau_0$, where $\Omega$ is the
perturbation frequency, and $\tau_0$ is the initial phase of the
perturbation. The quantities ${\cal F}$, ${\cal G}$, $a$, $b$ are
constant parameters.

The frequency of the pendulum small-amplitude oscillations is
given by

\vspace{-3mm}

\begin{equation}
\omega_0 = ({\cal F G})^{1/2} .
\label{ome0}
\end{equation}

\noindent The so-called adiabaticity parameter, measuring the
relative frequency of perturbation, is introduced as

\vspace{-3mm}

\begin{equation}
\lambda = \frac{\Omega}{\omega_0} , \label{lambda}
\end{equation}

\noindent see \cite{C79}.

In the well-known phase portrait ``$\phi$--$p$'' of the
non-perturbed pendulum, a single domain of librations, bounded by
the non-perturbed separatrix, is present. If the perturbations are
``switched on'' (i.e., $\varepsilon \neq 0$), a section of the
phase space of motion can be constructed, e.g., at $\tau = 0
\mbox{ mod } 2 \pi$. If the perturbation frequency is relatively
large, the separation of resonances in the momentum $p$ is large
and they almost do not interact. On reducing the frequency of
perturbation, the resonances approach each other and appreciable
chaotic layers emerge in the vicinity of the separatrices. On
reducing further on the perturbation frequency, the resonances
start to overlap, and these layers merge into a single chaotic
domain; see illustrations in \cite{S14PLA}.

The near-separatrix motion is conveniently described by separatrix
maps \cite{C79,S14PLA,S20}. They represent the motion in a
discrete way: the explored system's state is mapped discretely at
the time moments of its passage of positions of equilibria. For
FMR1, one iteration of the separatrix map corresponds to a
half-period of the model pendulum's libration or a period of its
rotation. Since the motion is near-separatrix, these periods can
be very large, up to infinity.

Let $w$ be a variable that denotes the relative (with respect to
the separatrix value) pendulum energy, $w \equiv (H_0 / {\cal F})
- 1$, and $\tau$ be the phase angle of perturbation. The
near-separatrix motion of the perturbed pendulum~(\ref{h}) with
asymmetric perturbation ($a \neq b$) is described by the so-called
separatrix algorithmic map \cite{S99}. In the symmetric case $a =
b$, the latter reduces to the ordinary separatrix map

\vspace{-3mm}

\begin{eqnarray}
& & w_{i+1} = w_i - W \sin \tau_i,  \nonumber \\
& & \tau_{i+1} = \tau_i +
                 \lambda \ln \frac{32}{\vert w_{i+1} \vert}
                 \ \ \ (\mbox{mod } 2 \pi),
\label{sm}
\end{eqnarray}

\noindent first written in this form in \cite{C79,C77}; an
expression for the parameter $W$, valid at any value of the
adiabaticity parameter $\lambda$ \cite{S00JETP,S98PS}, is

\vspace{-3mm}

\begin{equation}
W = \varepsilon \lambda \left( A_2(\lambda) + A_2(-\lambda)
\right) = 4 \pi \varepsilon \frac{\lambda^2}{\sinh \frac{\pi
\lambda}{2}}. \label{W}
\end{equation}

\noindent Here $\lambda$ is given by Eq.~(\ref{lambda}),
$\varepsilon = a / {\cal F}$, $\eta = b / a$. The Melnikov--Arnold
integral $A_2(\lambda)$, defined in \cite{C79}, at any value of
$\lambda$ (in the interval from zero to infinity) is given by the
formula

\vspace{-3mm}

\begin{equation}
A_2(\lambda) = 4 \pi \lambda \frac{ \exp (\pi \lambda / 2)}{ \sinh
(\pi \lambda)},
\end{equation}

\noindent see \cite{S00JETP,S98PS}.

Formula~(\ref{W}) differs from that given in \cite{C79,LL92} by
the addend $A_2(-\lambda)$, which is small if $\lambda \gg 1$.
However, its contribution is significant if $\lambda$ is small
\cite{S98PS}, i.e., in the case of adiabatic chaos.

An equivalent form of Eqs.~(\ref{sm}), used, e.g., in
\cite{CS84,S98PLA}, is

\vspace{-3mm}

\begin{eqnarray}
     y_{i+1} &=& y_i + \sin x_i, \nonumber \\
     x_{i+1} &=& x_i - \lambda \ln \vert y_{i+1} \vert + c
                   \ \ \ (\mbox{mod } 2 \pi),
\label{sm1}
\end{eqnarray}

\noindent where $y = w / W$, $x = \tau + \pi$; and the parameter

\vspace{-3mm}

\begin{equation}
c = \lambda \ln \frac{32}{\vert W \vert}. \label{c}
\end{equation}

It is important that the borderline $\lambda \approx 1/2$,
determined in \cite{S08PLA} to separate adiabatic chaos from
non-adiabatic one, does not separate the realm of resonance
overlap from that of non-overlap: the borderline lies much higher
in $\lambda$; e.g., in the phase space of the standard map, the
integer resonances start to overlap, on decreasing $\lambda$, at
the stochasticity parameter value $K = K_\mathrm{G} =
0.9716\ldots$ \cite{C79,Meiss92}, i.e., already at $\lambda = 2
\pi / K_\mathrm{G}^{1/2} \approx 6.37$.\footnote{A well-known
important constant of the standard map dynamics is the critical
value of $K$, namely, $K_\mathrm{G} = 0.971635406\ldots$; see,
e.~g., \cite{Meiss92}.}

\section{The second fundamental model of resonance}
\label{sect_FMR2}

For the second fundamental model of perturbed nonlinear resonance
(FMR2) we adopt the following Hamiltonian:

\vspace{-3mm}

\begin{equation}
H = \frac{{\cal G} (p - \delta)^2}{2} - {\cal F} p^{k/2} \cos k
\phi + a p^{k/2} \cos(k \phi - \tau) + b p^{k/2} \cos(k \phi +
\tau) , \label{h2k}
\end{equation}

\noindent where $\delta$ is the detuning parameter, and $k = 1, 2,
3, \dots$ (i.e., a natural number). The Hamiltonian's unperturbed
part

\begin{equation}
H_0 = \frac{{\cal G} (p - \delta)^2}{2} - {\cal F} p^{k/2} \cos k
\phi , \label{h2k0}
\end{equation}

\noindent can be obtained from the original definition of FMR2 in
\cite{HL83} by simple canonical transformations. The Hamiltonians
of kind~(\ref{h2k0}) are historically called the {\it Andoyer
Hamiltonians}; their general description in the $k=1$ case can be
found in Appendix~C in \cite{FM07}.

The coefficients of the cosine terms in~(\ref{h2k}) all contain
$p^{1/2}$; this is due to the D'Alembert rules for the Fourier
expansions of Hamiltonians in orbital dynamics; see section 1.9.3
in \cite{Morbi02}.

In what follows, we consider solely the case $k=1$, as the most
important one. Generalizations to $k=2, 3, 4, \dots$ are
straightforward. Therefore, we consider the Hamiltonian of the
form

\begin{equation}
H = \frac{{\cal G} (p - \delta)^2}{2} - {\cal F} p^{1/2} \cos \phi
+ a p^{1/2} \cos(\phi - \tau) + b p^{1/2} \cos(\phi + \tau) .
\label{h2}
\end{equation}

From analysis in \cite{HL83,FM07}, it follows that the separatrix
exists at $\delta \geq 0$; therefore, the separatrix splitting and
chaotic behavior can be observed at $\delta \geq 0$, and we limit
our study to this range of the detuning parameter values.

\begin{figure}
\begin{center}
\includegraphics[width=0.7\textwidth]{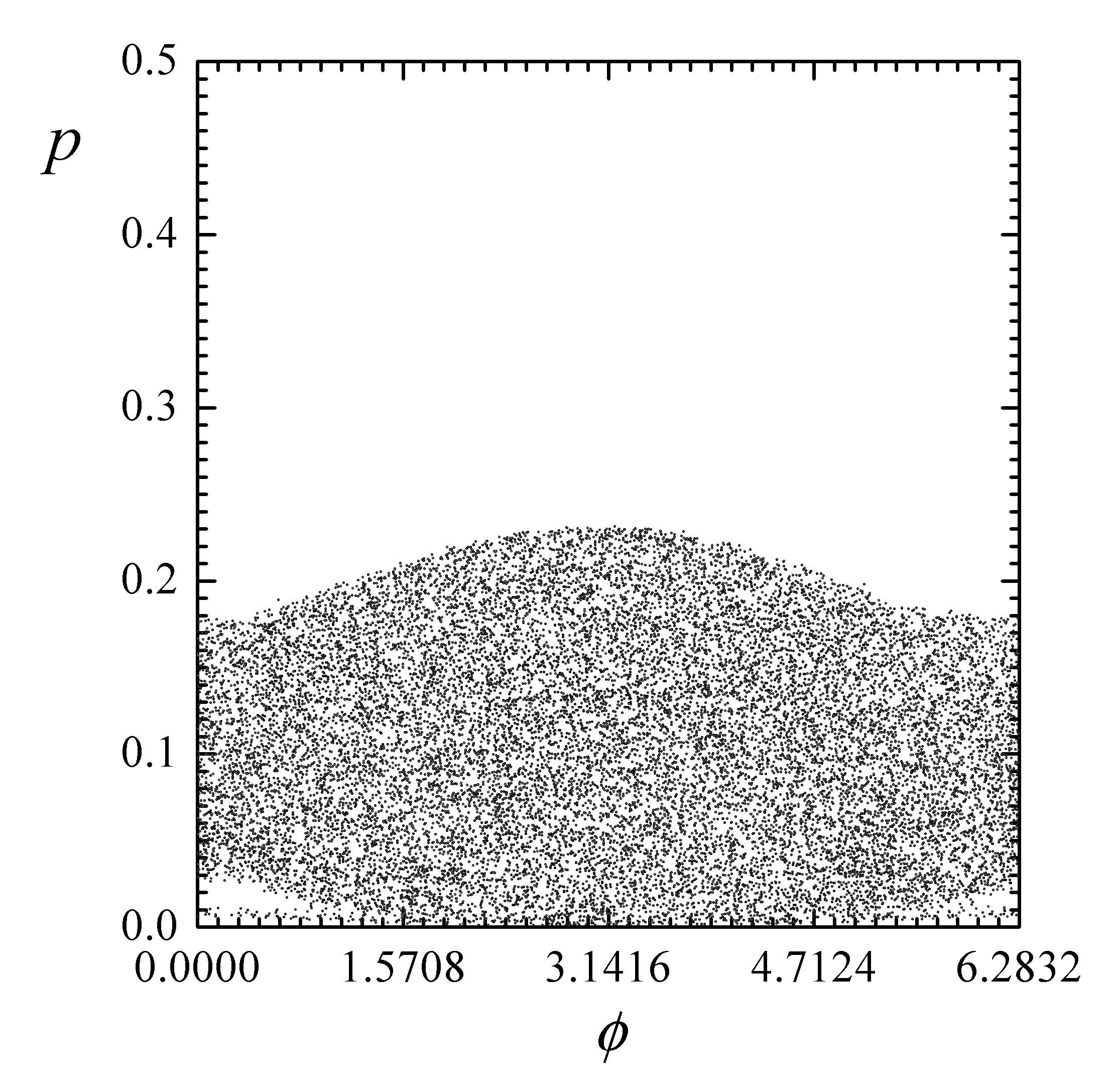}
\end{center}
\caption{\small A chaotic resonance triplet in FMR2. Strong
overlap ($\lambda = 1$), small detuning ($\delta = 0.1$).}
\label{fig1}
\end{figure}

\begin{figure}
\begin{center}
\includegraphics[width=0.7\textwidth]{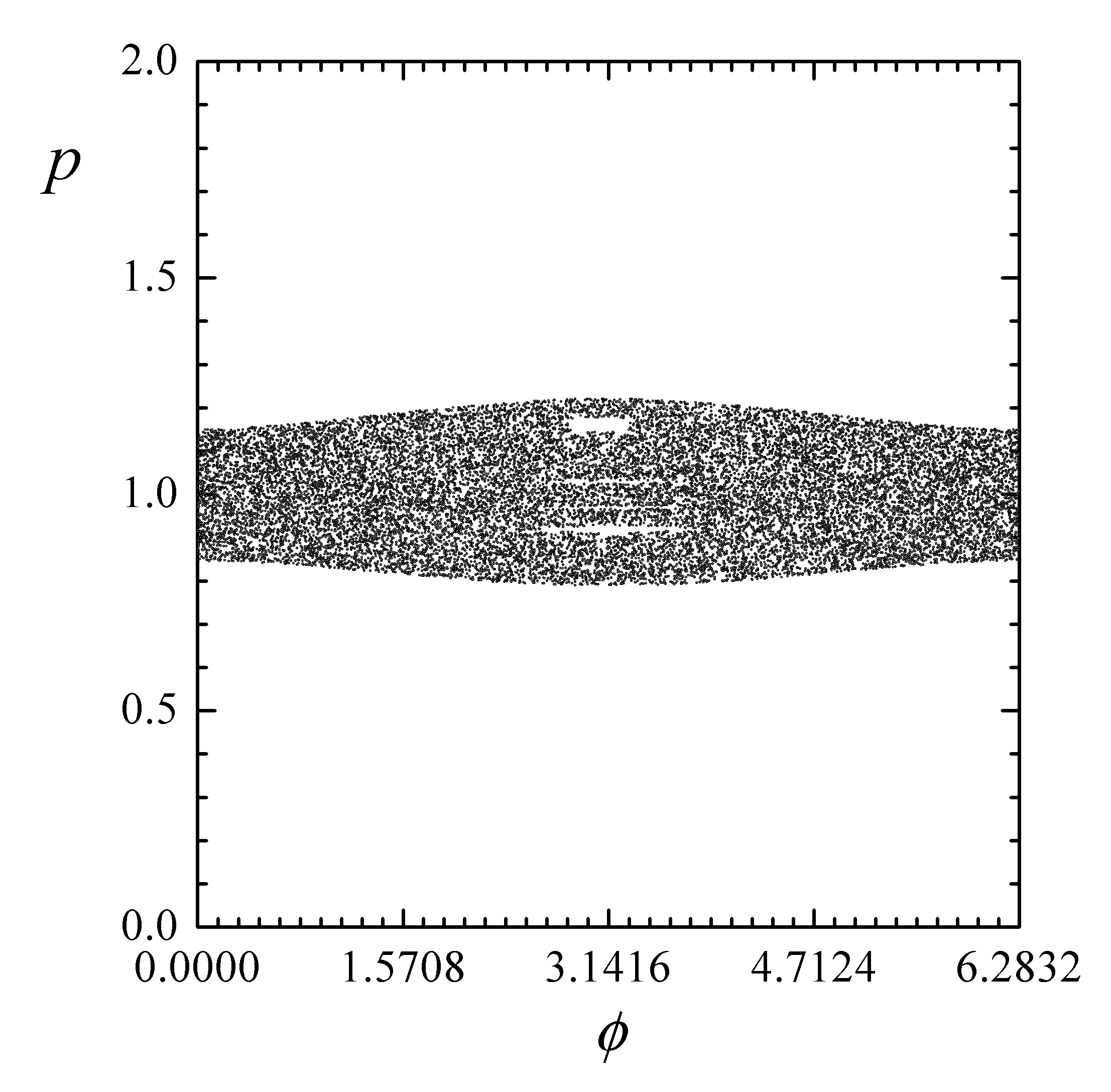}
\end{center}
\caption{\small Same as Fig.~\ref{fig1}, but for moderate detuning
($\delta = 1.0$).} \label{fig2}
\end{figure}

It is instructive to see how Poincar\'e sections for the perturbed
FMR2 look like; for an illustrative comparison with those for
FMR1, presented in \cite{S14PLA,S20}. Two representative examples
are shown in Figs.~\ref{fig1} and \ref{fig2}. For their
construction, we set $a=b$, $\varepsilon = a/{\cal F} =1$
(large-amplitude perturbation), and take two values (small and
moderate ones) of the detuning parameter, namely, $\delta = 0.1$
and 1.0. The phase space section is taken on the period of
perturbation, i.e., at $\tau = 0 \mod 2 \pi$. The numerical
integrator is the same as used extensively below for computations
in Section~\ref{lesfmr2}.

One may see that, in Fig.~\ref{fig1}, the Poincar\'e section for
the perturbed FMR2 looks qualitatively similar to those of the
perturbed FMR1 (presented, e.g., in \cite{S14PLA}), whereas in
Fig.~\ref{fig1}, where the detuning parameter $\delta$ is small,
it is rather specific.

\section{Lyapunov exponents in the first FMR}
\label{lesfmr1}

Let us briefly review known results on Lyapunov exponents in FMR1.
In \cite{S02CR}, an approach for estimating the maximum Lyapunov
exponent of the near-separatrix motion in FMR1 was proposed, based
on the separatrix map theory. In this approach, the maximum
Lyapunov exponent $L$ of the chaotic motion of system~(\ref{h}) is
represented as the ratio $L_\mathrm{sx} / T$, where
$L_\mathrm{sx}$ is the maximum Lyapunov exponent of the system's
separatrix map, and $T$ is the average period of rotation (or,
equivalently, the average half-period of libration) of the
resonance phase $\phi$ inside the chaotic layer. For convenience,
we introduce the non-dimensional quantity $T_\mathrm{sx} = \Omega
T$. Therefore,

\vspace{-3mm}

\begin{equation}
L = \Omega \frac{L_\mathrm{sx}}{T_\mathrm{sx}}. \label{lceh}
\end{equation}

By definition, the Lyapunov time $T_\mathrm{L} = L^{-1}$.

Here we briefly present analytical expressions for the Lyapunov
timescales $T_\mathrm{L}$ of motion in multiplets (namely,
triplets and supermultiplets) of interacting resonances in FMR1,
as derived in \cite{S14PLA,S20,S07IAU,S08MN} on the basis of the
separatrix map theory.

We set $a=b$ in Eq.~(\ref{h}), and assume $\lambda > 1/2$
(non-adiabatic chaos). Then one has a symmetric triplet of
interacting resonances, and chaos is non-adiabatic. Following
\cite{S02CR}, we take the $\lambda$ dependence of the maximum
Lyapunov exponent of the separatrix map~(\ref{sm1}) in the form

\vspace{-3mm}

\begin{equation}
L_\mathrm{sx}(\lambda) \approx C_h \frac{2 \lambda}{1 + 2
\lambda}, \label{Lsx}
\end{equation}

\noindent where $C_h \approx 0.80$ is Chirikov's constant
\cite{S04JETPL}. The average increment of $\tau$ (proportional to
the average libration half-period, or the average rotation period)
inside the chaotic layer is \cite{C79,S02CR}:

\vspace{-3mm}

\begin{equation}
T_\mathrm{sx}(\lambda, W) \approx \lambda \ln \frac{32 e}{\lambda
| W |}, \label{Tsx}
\end{equation}

\noindent where $e$ is the base of natural logarithms.

Then, the Lyapunov time is given by \cite{S14PLA,S07IAU}:

\vspace{-3mm}

\begin{equation}
T_\mathrm{L} = \frac{T_\mathrm{pert}}{2 \pi}
\frac{T_\mathrm{sx}}{L_\mathrm{sx}} \approx
T_\mathrm{pert} \frac{1 + 2 \lambda}{4 \pi C_h}
\ln \frac{32 e}{\lambda | W |},
\label{TLft}
\end{equation}

\noindent where $T_\mathrm{pert} = 2 \pi / \Omega$ is the
perturbation period.

If $\lambda < 1/2$ (adiabatic chaos), the diffusion in canonical
momentum across the chaotic layer is slow, and on a short time
interval the trajectory of the separatrix map~(\ref{sm1}) follows
close to some current ``guiding'' curve; this observation allows
one to deduce the Lyapunov timescale as \cite{S07IAU,S08MN}:

\vspace{-3mm}

\begin{equation}
T_\mathrm{L} \approx \frac{T_\mathrm{pert}}{2 \pi} \left( \ln
\left| 4 \sin \frac{c}{2} \right| + \frac{c}{\lambda} \right) ,
\label{TLst1}
\end{equation}

\noindent where $c = \lambda \ln \frac {32}{|W|}$, as given by
formula~(\ref{c}). Formula~(\ref{TLst1}) is valid when the
parameter $c$ (approximately equal to $\lambda \ln
\frac{4}{\lambda | \varepsilon |}$ in the adiabatic case) is not
too close to $0 \mbox{ mod } 2 \pi$; see \cite{S08MN} for details.

If $\lambda \ll 1$, then $W \approx 8 \varepsilon \lambda$, and

\vspace{-3mm}

\begin{equation}
T_\mathrm{L} \approx \frac{T_\mathrm{pert}}{2 \pi} \ln \left|
\frac{16}{\varepsilon \lambda} \sin \left( \frac{\lambda}{2} \ln
\frac{4}{ |\varepsilon| \lambda} \right) \right|.
\label{TLst}
\end{equation}

Let the number of resonances in a resonance multiplet in FMR1 be
greater than three. In applications, especially concerning
three-body mean-motion resonances, this number can be very large
\cite{Morbi02}; then, one has a supermultiplet. To model it, we
consider infinitely many interacting equally-sized equally-spaced
resonances, and the motion is described by the standard map

\vspace{-6mm}

\begin{eqnarray}
y_{i+1} &=& y_i + K \sin x_i \ \ \ (\mbox{mod } 2 \pi), \nonumber \\
x_{i+1} &=& x_i + y_{i+1} \ \ \ (\mbox{mod } 2 \pi) .
\label{stm2}
\end{eqnarray}

\noindent Indeed, its Hamiltonian is \cite{C79}:

\vspace{-3mm}

\begin{equation}
H = \frac{y^2}{2} + \frac{K}{(2 \pi)^2} \sum_{k=-N}^N \cos(x - k t) , \label{h_stm2}
\end{equation}

\noindent where $N = \infty$. The variables $x_i$, $y_i$ of
map~(\ref{stm2}) correspond to the variables $x(t_i)$, $y(t_i)$ of
the continuous system~(\ref{h_stm2}) taken stroboscopically at
time moduli $2 \pi$; see, e.g., \cite{C79}.

The asymptotic formula for the maximum Lyapunov exponent of the
standard map at $K \gg 1$ was derived in \cite{C79}: $L(K) \propto
\ln \frac{K}{2}$. For any $K$ in the full range $K \in (0,
\infty)$, a suitable expression for the Lyapunov time was proposed
in \cite{S04JETPL,S04PLA}:

\vspace{-3mm}

\begin{equation}
T_\mathrm{L} \approx T_\mathrm{pert} \cdot
\begin{cases}
\displaystyle{\frac{7.50}{K}} (\approx 0.190 \lambda^2), &
\text{if $K < 1.1$ (or, if $\lambda > 6.0$)}, \\
2.133 (K - 1.037)^{-1/2} , &
\text{if $1.1 \leq K < 4.4$ (or, if $3.0 < \lambda \leq 6.0$)}, \\
\displaystyle{\left( \ln \frac{K}{2} + \frac{1}{K^2} \right)^{-1}}
, & \text{if $K \geq 4.4$ (or, if $\lambda \leq 3.0$)},
\end{cases}
\label{TLinf}
\end{equation}

\noindent where

\vspace{-3mm}

\begin{equation}
K = (2 \pi /\lambda)^2 .
\label{Kla}
\end{equation}

\noindent At $K \lesssim 1$, the expression~(\ref{TLinf}) was
derived from the separatrix map theory; and at $1 \lesssim K
\lesssim 4$, a fitting formula was picked up to sew together two
asymptotic regimes; see \cite{S04JETPL,S04PLA}.

Assuming that the limiting case $M=\infty$ for the number of
resonances in a multiplet describes the situation at $M \gg 1$,
formula~(\ref{TLinf}) can be applied to estimate the Lyapunov
timescale when the number of interacting resonances is large.

Given $T_\mathrm{L}$, the corresponding maximum Lyapunov exponent
$L$ can always be calculated as $L = 1 / T_\mathrm{L}$. Also note
that, everywhere in the presented above formulas (\ref{TLft},
\ref{TLst1}, \ref{TLst}, \ref{TLinf}) for $T_\mathrm{L}$, the
proportionality of $T_\mathrm{L}$ to $T_\mathrm{pert}$ can always
be changed to proportionality to $T_0 = 2 \pi / \omega_0$ by
applying the substitution $T_\mathrm{pert} = T_0 / \lambda$.
Conversely, the Lyapunov exponent $L$ can be formally written down
as proportional either to the perturbation frequency $\Omega$, or
to the frequency of small-amplitude resonance phase libration
$\omega_0 = \Omega / \lambda$.

\section{Lyapunov exponents in the second FMR}
\label{lesfmr2}

By a constant shift by $\delta$ in the canonical momentum variable
$p$, the Hamiltonian~(\ref{h2}) of the second fundamental model of
perturbed nonlinear resonance can be readily reduced to the form

\vspace{-3mm}

\begin{equation}
H = \frac{{\cal G} p^2}{2} - {\cal F} (p+\delta)^{1/2} \cos \phi
+ a (p+\delta)^{1/2} \cos(\phi - \tau) + b (p+\delta)^{1/2}
\cos(\phi + \tau) , \label{h2sh}
\end{equation}

\noindent where the henceforth used shifted momentum is denoted,
for convenience, by the same letter $p$ as the original unshifted
one.

As already adopted above, we consider the case of a symmetric
triplet, $a=b$. Assuming that the shifted momentum $p \ll \delta$,
we expand the coefficients of the trigonometric terms in power
series in $p/\delta$, keeping only the first-order terms:

\begin{equation}
H \cong \frac{{\cal G} p^2}{2} -
{\cal F} \delta^{1/2} \left( 1 + \frac{p}{2 \delta} \right) \cos \phi +
a \left( 1 + \frac{p}{2 \delta} \right)
[ \cos(\phi - \tau) + \cos(\phi + \tau) ] . \label{h2ps}
\end{equation}

\noindent Thus, in this approximation, which should be valid at
sufficiently large detuning $\delta$, FMR2 is reduced to FMR1, but
the small-amplitude phase oscillation frequency is now given by

\begin{equation}
\omega_0 = ({\cal F G})^{1/2} \delta^{1/4}. \label{ome0ps}
\end{equation}

\noindent The adiabaticity parameter is defined as usual, $\lambda
= \Omega / \omega_0$, but now $\omega_0$ is given by
formula~(\ref{ome0ps}). The relative amplitude $\varepsilon$ of
perturbation keeps the same form as in FMR1: $\varepsilon =
a/{\cal F}$.

On the unperturbed separatrix, if ${\cal G}=1$, one has $p = 2
\omega_0 \cos \frac{\phi}{2}$ (e.g., \cite{C79}), therefore, the
approximation~(\ref{h2ps}) is valid at $\delta \gg ({\cal F /
G})^{2/3}$.

In this approximation, one may use the separatrix map theory and
estimate the Lyapunov timescales and exponents by applying the
same formulas as in the FMR1 case (using formulas (\ref{TLft},
\ref{TLst1}, \ref{TLst}, \ref{TLinf}); only the quantities
$\omega_0$ and $\lambda$ are calculated now taking into account
the detuning parameter $\delta$, by using formula~(\ref{ome0ps}).

Let us see how this correspondence expresses itself in numerical
experiments. To compute the Lyapunov exponents and timescales we
use numerical algorithms and software developed in
\cite{SK02,KS05}. The programs allow one to calculate the full
Lyapunov spectra, by employing the HQRB numerical method
\cite{BUP97}, based on the QR decomposition of the tangent map
matrix. (The decomposition uses the Householder transformation,
hence the abbreviation ``HQRB''.)

For computations of orbits in the FMR1 and FMR2 models we use the
integrator by Hairer et al.\ \cite{HNW87}; it realizes an explicit
8th order Runge--Kutta method due to Dormand and Prince, with the
step size control.

\begin{figure}[th!]
\centering
\includegraphics[width=0.7\textwidth]{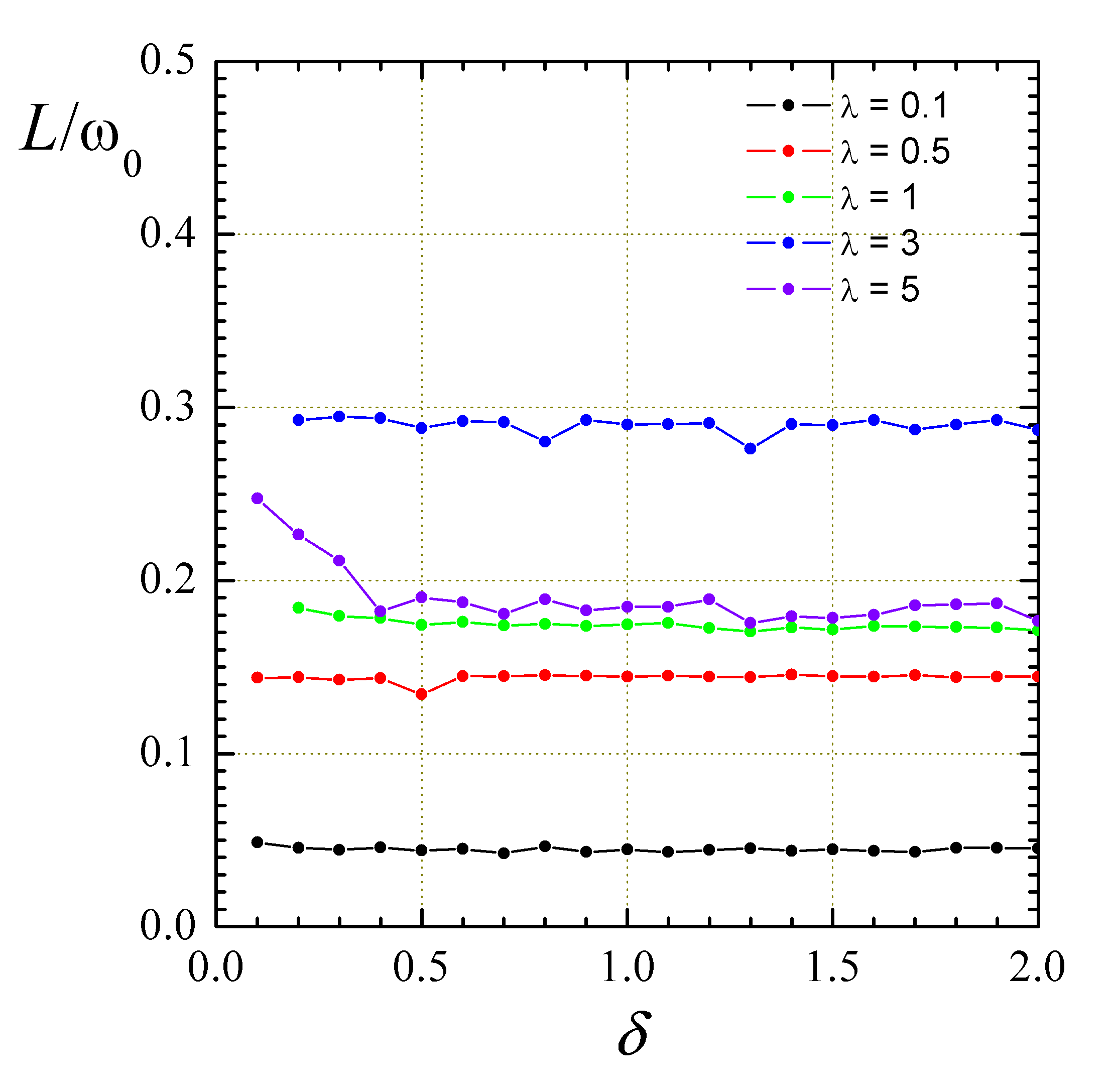}
\caption{The $\delta$ dependences of the $\omega_0$-normalized
maximum Lyapunov exponent $L$, as obtained in our numerical
experiments for triplets of equally-sized equally-spaced
resonances in FMR2, at several values of the adiabaticity
parameter $\lambda$.} \label{fig3}
\end{figure}

The $\delta$ dependences of the $\omega_0$-normalized maximum
Lyapunov exponent $L$ for triplets of equally-sized ($\varepsilon
= 1$) equally-spaced resonances in FMR2 are shown in
Fig.~\ref{fig3}, as obtained in our numerical experiments at
several values of the adiabaticity parameter $\lambda$. The
approximate constancy of $L/\omega_0$ with respect to variation of
$\delta$ in the given range from 0.1 up to 2.0 is evident. Since,
in the whole explored range of $\delta$, the validity condition
$\delta > ({\cal F / G})^{2/3}$ holds, this is just what has been
expected on the basis of our theory. Note that the approximate
constancy is observed even at $\delta = 0.1$, when the inequality
in the validity condition approaches equality.

\section{Dependences of Lyapunov exponents on the adiabaticity parameter}
\label{sect_LERM}

The Hamiltonian of a multiplet of $2 M + 1$ (where $M=1, 2, 3,
\dots$) equally-spaced resonances in FMR1 is given by
\cite{S14PLA}:

\vspace{-3mm}

\begin{equation}
H = {{{\cal G} p^2} \over 2} - {\cal F} \cos \phi +
    \sum_{k=1}^{M} a_k \cos(\phi - k \tau) +
    \sum_{k=1}^{M} b_k \cos(\phi + k \tau).
\label{hmulti}
\end{equation}

\noindent We set $\varepsilon_k = a_k / {\cal F} = b_k / {\cal F}
= 1$, i.e., the resonances in the multiplet are equally-sized.

Note that $L/\omega_0$ depends somewhat on the perturbation
amplitude $\varepsilon$, however, FMR1 systems with $\varepsilon =
0.01$ and with $\varepsilon = 1$ (i.e., differing in the amplitude
a hundred times) differ in the Lyapunov exponent only by about
three times \cite{S14PLA}. Systems with $\varepsilon \sim 1$
prevail in applications; thus, the dependence on $\varepsilon$ can
be usually ignored, and the case $\varepsilon \sim 1$ is therefore
representative.

We compute the maximum Lyapunov exponents for a number of systems
with Hamiltonian~(\ref{hmulti}) with various $M$, and for a
doublet case (where $a_k = {\cal F}$ and $b_k = 0$). To this end,
we use the algorithms and software already used above in
constructing Fig.~\ref{fig3}. For the infinitet, i.e., for the
standard map, analytical formulas~(\ref{TLinf}) giving
$T_\mathrm{L}$ are used, and $L = 1 / T_\mathrm{L}$.

\begin{figure}[th!]
\centering
\includegraphics[width=0.84\textwidth]{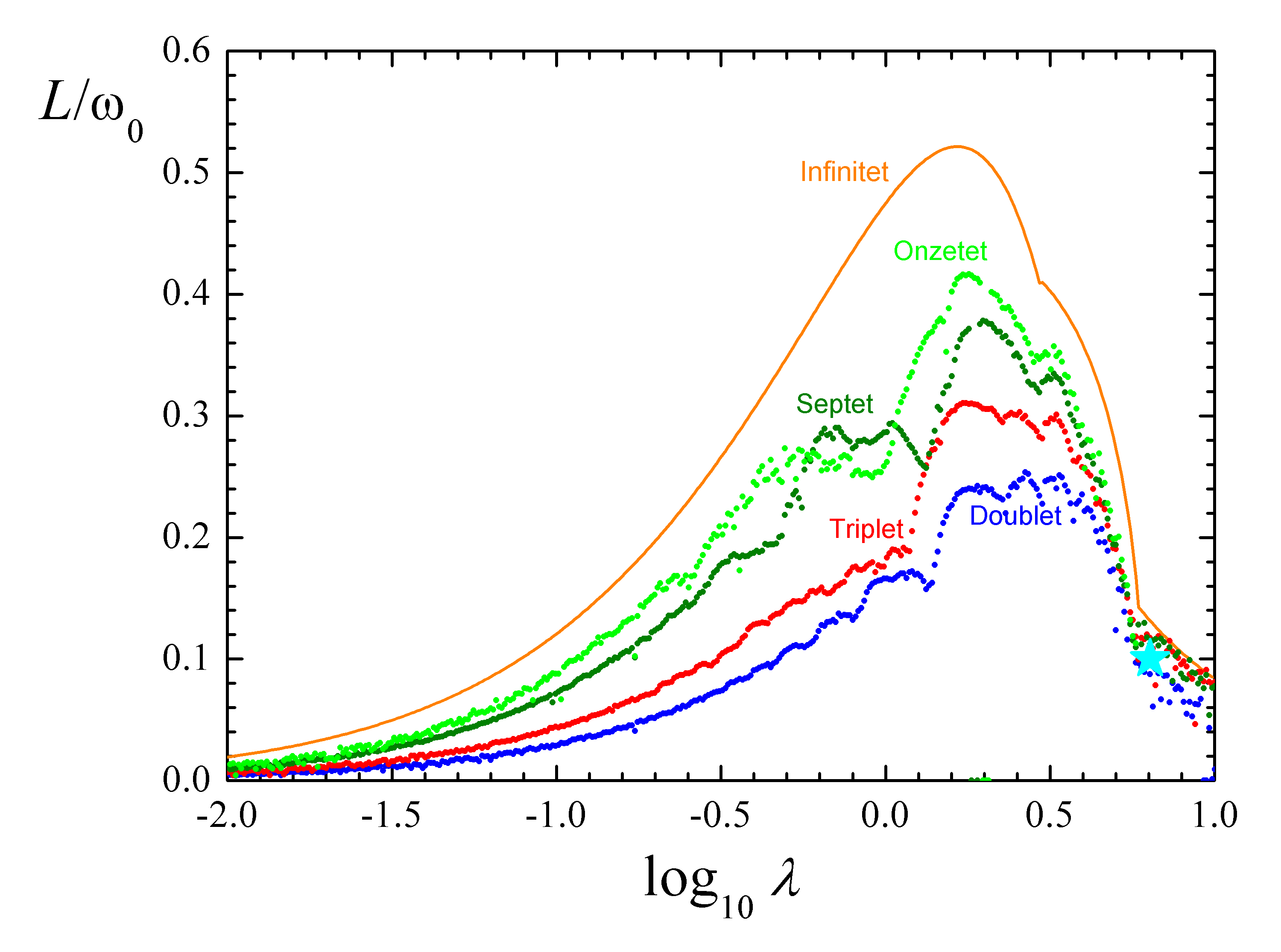}
\caption{The $\lambda$ dependences of the $\omega_0$-normalized
maximum Lyapunov exponent $L$, for multiplets of equally-sized
equally-spaced resonances in FMR1, as obtained in our numerical
experiments. The upper solid curve represents the standard map
theory for the infinitet. For FMR2, the dependences are the same,
once the detuning $\delta$ is large enough. The location
corresponding to the marginal overlap of resonances in the
multiplets is shown by a cyan star.} \label{fig4}
\end{figure}

The obtained numerical-experimental $\lambda$ dependences of the
maximum Lyapunov exponent (normalized by $\omega_0$) for several
multiplets of equally-sized equally-spaced resonances are shown in
Fig.~\ref{fig4}. The upper solid curve, given by
Eqs.~(\ref{TLinf}), represents the standard map theory for the
infinitet.

One may see that the $\lambda$ dependences for the triplet
($M=1$), septet ($M=3$, 7 interacting resonances) and onzetet
($M=5$, 11 interacting resonances) occupy in Fig.~\ref{fig4}
intermediate (in the vertical axis) positions between the
dependences for the doublet and the infinitet. Thus, as already
stated in \cite{S14PLA}, the value of $L/\omega_0$ in the
multiplet of equally-spaced equally-sized resonances is minimum in
the doublet case and maximum in the infinitet case. The upper
solid curve, giving $L/\omega_0$ for the standard map, thus
provides an upper limit for the possible $L/\omega_0$ values in
resonance multiplets.

As follows from our results in Section~\ref{lesfmr2}, the
dependences of $L/\omega_0$ on $\lambda$ for FMR2 should be
practically the same as for FMR1, once the detuning $\delta$ is
large enough. Therefore, Fig.~\ref{fig4}, if recomputed for FMR2
instead of FMR1, would be the same, at any relevant choice of
$\delta$.

As already noted above, the resonances in the infinitet start to
overlap, on decreasing $\lambda$, at $K = K_\mathrm{G} \approx
0.9716$, i.e., at $\lambda = 2 \pi / K_\mathrm{G}^{1/2} \approx
6.37$; see formula (\ref{Kla}). Therefore, the range in $\lambda$
in Fig.~\ref{fig4} corresponds to the overlap condition except at
$\lambda \gtrsim 6.4$, i.e., at $\log_{10} \lambda \gtrsim 0.8$.

Moreover, according to Fig.~\ref{fig4}, the critical value of
$\log_{10} \lambda \sim 0.8$, at which the resonances in the
multiplet start to overlap, is approximately the same for all
multiplets. Indeed, the dependences for all multiplets converge at
this location, giving one and the same $L/\omega_0 \sim 0.1$; this
is natural, because at high values of $\lambda$ the effect of
perturbing harmonics is exponentially small with increasing the
$k$ index in Eq.~(\ref{hmulti}).

The location corresponding to the marginal overlap (the
stochasticity parameter $K = (2 \pi /\lambda)^2 \sim 1$) of
resonances in the multiplets is shown by a cyan star.

\section{Lyapunov timescales in the Solar system}
\label{sect_As}

Now let us see how the developed theory can be used to estimate
Lyapunov timescales in the Solar planetary system.

For the inner Solar system, we consider the system model proposed
in \cite{BMH15}, and for the outer one that proposed in
\cite{MH99}. Mercury's chaotic dynamics is represented in the
first model, and that of three outer giant planets in the second
one.

First of all, it is important to outline that the chaotic behavior
in the both models is attributed in \cite{BMH15} and \cite{MH99}
as due to a {\it marginal} overlap of interacting resonances: in
the both cases the stochasticity parameter $K_\mathrm{eff}$ is
estimated as $\sim 1$.

Therefore, in the diagram in Fig.~\ref{fig4}, the system in each
of the models should be placed on appropriate $L / \omega_0$
curves at $\lambda \sim 6$. At such a high value of $\lambda$, the
curves converge, thus the system's vertical location is determined
rather unambiguously. In Fig.~\ref{fig4}, this expected position
is already marked by the cyan star. At the given location, one has
$L \sim 0.1 \omega_0$.

Now let us consider the inner and outer models separately in more
detail.

\subsection{Inner Solar system}

In extensive numerical experiments
\cite{ML21AA,SW92,RT18,HML22,La89,La90} it was established that
the planetary orbits in the inner Solar System possess Lyapunov
times $\sim$5--7 million years. Several analytical estimates also
exist. In \cite{BMH15}, the Lyapunov time was estimated using a
formula of the form $T_L = \pi / \langle \omega_0 \rangle$ (in our
terms), where $\langle \omega_0 \rangle$ is the average of the
unstable eigenvalues (equal to the frequencies of small-amplitude
libration on resonances) of the overlapping resonances responsible
for chaos observed; an estimate $T_L \sim 1.4$ million years was
thus obtained \cite{BMH15}.

How this estimate conforms to our theory? According to
Fig.~\ref{fig4}, at $\lambda \approx 6$ (i.e., at $K \approx 1$),
one has $L/\omega_0 \sim 0.1$. Therefore, the coefficient $\pi$ in
the formula $T_L = \pi / \langle \omega_0 \rangle$ in \cite{BMH15}
should be replaced by a factor $\approx 10$. Then, one has $T_L
\sim 5$~Myr, instead of $T_L \sim 1.4$~Myr. The value of 5~Myr
agrees much better with known massive numerical-experimental
estimates, obtained and cited in \cite{ML21AA,ML22AA}.

Note that, generally, when the Lyapunov timescales are estimated
as proportional to the inverses of the unstable eigenvalues, a
coefficient of $2 \pi$ is often introduced \cite{BMH15,MH99,MH01},
making the resulting formula similar to that for the period of
small-amplitude librations about stable equilibria; this procedure
is equivalent to taking $L/\omega_0 = 1$ and $T_\mathrm{L} = 2 \pi
/ L$. However, the latter formula is purely heuristic, and has no
analogue in numerical-experimental determinations of Lyapunov
times, where the Lyapunov time is determined as simply the inverse
of the maximum Lyapunov exponent $L$, without multiplying this
inverse by $2 \pi$.

Apart from the analytical one, the numerical-experimental
$T_\mathrm{L}$ value obtained in \cite{BMH15} seems to be also
underestimated by a factor of 3--5; a possible cause is that the
MEGNO technique used in \cite{BMH15} provides only local values of
Lyapunov exponents, when their computation times are not large
enough.

On the grounds that the value of $T_\mathrm{L}$ presented in
\cite{BMH15} is too small, in contradiction with massive
numerical-experimental data, the inner Solar system model by
\cite{BMH15} may seem to be oversimplified, as argued in
\cite{ML21AA,ML22AA}. Conversely, we conclude that this model
actually does provide a reasonable estimate $T_\mathrm{L} \sim
5$~mln years (several times greater than that derived in
\cite{BMH15} on heuristic grounds), if the developed above
analytical approach, instead of heuristic estimates, is used.
Within this approach, the model conforms with known massive
numerical-experimental estimations of $T_\mathrm{L}$ in full
problem settings.

\subsection{Outer Solar system}

In \cite{MH99}, the origin of chaos in the outer Solar system was
attributed to interaction of subresonances in a multiplet
corresponding to a particular Jupiter--Saturn--Uranus three-body
mean-motion resonance. Our Solar system, as it is well known, is
close to the 5/2 Jupiter--Saturn two-body mean-motion resonance,
and it is also not far from the 7/1 Jupiter--Uranus two-body
mean-motion resonance. Neither of the corresponding resonant
arguments librate, i.e., the system is out of the both resonances,
although close to them. However, as established in \cite{MH99}, a
three-body linear combination of the resonance phases of these two
resonances may librate; namely, the 3J-5S-U7 three-body resonance

\vspace{-3mm}

\begin{equation}
3 \dot \lambda_\mathrm{J} - 5 \dot \lambda_\mathrm{S} - 7 \dot
\lambda_\mathrm{U} \sim 0 \label{res357}
\end{equation}

\noindent may actually be present. Here $\lambda_\mathrm{J}$,
$\lambda_\mathrm{S}$, and $\lambda_\mathrm{U}$ are the mean
longitudes of Jupiter, Saturn, and Uranus, respectively;
$\dot\lambda_\mathrm{J}$, $\dot\lambda_\mathrm{S}$, and
$\dot\lambda_\mathrm{U}$ are the corresponding mean motions (the
upper dots denote time derivatives).

The three-body resonance~(\ref{res357}) possesses a lot of
eccentricity-type, inclination-type, and
eccentricity-inclination-type subresonances, obeying the
D'Alembert rules. Their interaction presumably causes the observed
chaotic behaviour.

Only a marginal overlap of the subresonances was established in
\cite{MH99}; therefore, one has the stochasticity parameter $K
\sim 1$, and the adiabaticity parameter $\lambda = 2 \pi / K^{1/2}
\sim 6$. Therefore, the chaotic motion takes place in a strongly
non-adiabatic regime; chaos is non-adiabatic. In this respect, the
situation is the same as in the inner system, considered in the
previous Subsection.

In \cite{MH99}, the Lyapunov time was estimated as the period of
small-amplitude phase oscillations on the guiding subresonance in
the three-body resonance multiplet. As already discussed above,
the factor $2 \pi$ in the formula $T_\mathrm{L} = 2 \pi /\omega_0$
is generally irrelevant. If the period of a pulsating separatrix
is implied here (as in \cite{BMH15,MH97}), not that of small phase
oscillations, one should take care that the studied chaos is
adiabatic. Any approach for estimations of Lyapunov exponents
based on considering the slowly pulsating separatrix should be
used solely in the case of adiabatic chaos, i.e., at the
adiabaticity parameter values $\lambda \leq 1/2$; see \cite{S20}
and references therein. Extrapolating any formulas valid in the
adiabatic case to the non-adiabatic realm would result in serious
over-estimating the Lyapunov exponents and under-estimating the
Lyapunov timescales.

In the case of marginal resonance overlap, as already considered
in the previous Subsection, $\lambda \sim 2 \pi /
K_\mathrm{G}^{1/2} \sim 6$; then, from Fig.~\ref{fig4} we see that
the Lyapunov exponent $L$ itself is about ten times less than
$\omega_0$.

In \cite{MH99}, introducing the heuristic $2 \pi$ factor in a
formula for $T_\mathrm{L}$, i.e., taking $T_\mathrm{L} \sim 2 \pi
/ \omega_0$ instead of actual $T_\mathrm{L} = 1 / L \sim 1 /
\omega_0$), compensates for the heuristic adopting $L \sim
\omega_0$ instead of actual $L \sim 0.1 \omega_0$, at $\lambda
\approx 6$. The resulting estimate for $T_\mathrm{L}$, obtained in
\cite{MH99} as equal to $\sim 10^7$~yr, is therefore in a
satisfactory agreement with numerical-experimental data. Replacing
the heuristic factor $2 \pi$ by 10 (consistent with the diagram in
Fig.~\ref{fig4}) does not, however, effect the accuracy of the
$T_\mathrm{L}$ estimate much, since $\omega_0$ itself is known not
precisely enough.

\section{Conclusions}
\label{concl}

In this article, the problem of analytical estimation of the
Lyapunov exponents and Lyapunov timescales of the motion in
multiplets of interacting nonlinear resonances has been
considered.

To this end, we have elaborated a unified framework, based on the
separatrix map theory, which incorporates both an earlier approach
for the first fundamental model of perturbed resonance (given by
the perturbed pendulum Hamiltonian) and a new one for its second
fundamental model (given by the perturbed Andoyer Hamiltonian).

Within this framework, we have presented and discussed new
accurate estimates for the Lyapunov timescales of the inner and
outer subsystems of the Solar planetary system.

In particular, we have shown that the low-dimensional model of
\cite{BMH15} provides adequate estimates of the Lyapunov time for
the inner Solar system, which conforms with the known massive
numerical-experimental estimates in full problem settings.

For the outer Solar system, the presented theory, as applied
within the model \cite{MH99} for the 3J-5S-U7 three-body
mean-motion resonance, also provides adequate analytical
$T_\mathrm{L}$ values, in agreement with known
numerical-experimental estimates.

\section*{Funding}

This work was supported by ongoing institutional funding. No
additional grants to carry out or direct this particular research
were obtained.

\section*{Conflict of interest}

The author of this work declares that he has no conflicts of
interest.

\end{document}